\documentclass[aps,prl,floatfix,twocolumn]{revtex4-1}
\usepackage{amsmath,amssymb}
\usepackage{graphicx}
\usepackage{epsfig}
\usepackage[usenames]{color}

\usepackage{ulem}

\begin{document}

%\title{Valley photogalvanic effect of \textcolor{red}{photon-dressed} carriers in 2D Dirac materials}

\title{Photogalvanic currents in dynamically gapped Dirac materials}

%\title{Giant photogalvanic currents in dynamically gapped Dirac materials}

\author{V.~M.~Kovalev and I.~G.~Savenko}
%\email{vadimkovalev@isp.nsc.ru}
\affiliation{A.V.~Rzhanov Institute of Semiconductor Physics, Siberian Branch of Russian Academy of Sciences, Novosibirsk 630090, Russia\\
Center for Theoretical Physics of Complex Systems, Institute for Basic Science (IBS), Daejeon 34126, Korea}

%\author{I.~G.~Savenko}
%\affiliation{Center for Theoretical Physics of Complex Systems, Institute for Basic Science (IBS), Daejeon 34126, Korea}
%\affiliation{Basic Science Program, Korea University of Science and Technology (UST), Daejeon 34113, Korea}

\date{\today}

\begin{abstract}
We develop a microscopic theory of an unconventional photogalvanic effect in two-dimensional materials with the Dirac energy spectrum of the carriers of charge under strong driving. As a test bed, we consider a layer of a transition metal dichalcogenide, exposed to two different electromagnetic fields.
The first \textit{pumping} field is circularly-polarized, and its frequency exceeds the material bandgap. It creates an extremely nonequilibrium distribution of electrons and holes in one valley (K) and opens dynamical gaps, whereas the other valley (K$^\prime$) remains empty due to the valley-dependent interband selection rules.
The second \textit{probe} field has the frequency much smaller than the material bandgap. It generates intraband perturbations of the nonequilibrium carriers density, resulting in the photogalvanic current due to the trigonal asymmetry of the dispersions. 
This current shows thresholdlike behavior due to the dynamical gap opening and  renormalizations of the density of states and velocity of quasiparticles.
\end{abstract}

%\pacs{}

\maketitle

%---------------------------
%---------------------------
%---------------------------

\section{I. Introduction}
Two-dimensional (2D) quantum systems exposed to external powerful high-frequency electromagnetic (EM) fields exhibit a variety of fascinating phenomena~\cite{Ivchenko}, including dissipation-free electron transport~\cite{KibisPRL}, quasi-condensation~\cite{ExPol1,ExPol2}, and the photon drag effect~\cite{Wieck, Glazov, Entin, RefRPQ} among others.
%RefOurJETP, RefPDE
%
In the case of nearly-resonant excitation of a solid-state system and strong light-matter interaction, it is convenient to work with hybrid  photon-dressed quasiparticles, characterized by  nonequilibrium steady-state distribution functions~\cite{OurNJP}.
Their spectrum possesses a dynamical gap~\cite{OurNJP, ShelykhGap, Machlin}, determined by the amplitude of the external EM field~\cite{BlochOsc}.

Initially, dynamical gaps were studied in gapless materials such as graphene~\cite{Geim, RefSyzranov1,   RefSyzranov2}. Currently, valley physics of 2D materials~\cite{xiao2007valley}, in particular transition metal dichalcogenides (TMDs)~\cite{jung2015origin, xu2014spin}, is in focus~\cite{yao2008valley}. Their Brillouin zone contains two valleys K and K$^\prime$, coupled by the time reversal symmetry. Therefore in addition to momentum and spin, 2D semiconductors possess another degree of freedom, which refers to particular valley. It is especially practical, that spectrum of TMDs has large gap (e.g., in MoS$_2$ it amounts to $1.66\,\mathrm{eV}$~\cite{xiao2012coupled}), giving a possibility to study valley-resolved physics~\cite{RefMak, RefUbrig}.

They have symmetry properties similar to monolayer graphene with staggered sublattice potential. Due to the spatial inversion symmetry breaking, there occur transport effects described by a third-order generalized conductivity tensor.
A typical example is the photogalvanic effect (PGE) also called the photovoltaic effect, where the components of photoinduced current $j^\alpha$ are coupled with the components of the vector potential of an external EM field $A^\beta$ by the relation
\begin{equation}
j^\alpha=\chi_{\alpha\beta\gamma}A^\beta A^\gamma,
\end{equation}
where $\alpha,~\beta,~\gamma=x,~y,~z$ and $\chi_{\alpha\beta\gamma}$ is the photogalvanic third-order tensor, which can only exist (be finite) in noncentrosymmetric materials.

The microscopic origin of the conventional PGE is in the asymmetry of the interaction potential or the crystal-induced Bloch wave function~\cite{ivchenko, belinicher}.
It can also take place in 2D materials. However, there can appear an unconventional PGE due to the trigonal warping of the valleys.
This asymmetry of the particle dispersion leads to such fascinating phenomena as the second harmonic generation~\cite{GT}, purely valley currents~\cite{MEGT} and alignment of the photoexcited carriers in gapless materials~\cite{Portnoi1, Portnoi2}. Recently, the PGE produced by a weak EM fields and the spectrum warping of the valleys have also been discussed~\cite{MagarillEntinKovalev}.

The PGE currents in valleys K and K$^\prime$ flow in opposite directions. Consequently, the net current is zero due to the time reversal symmetry. To launch a nonzero current in such circumstances, one has to break the time reversal symmetry.
It can be done by an external electromagnetic field with circular polarization.
Indeed, the specific property of TMD materials is that they possess the polarization-sensitive interband optical selection rules: electrons in the valley K (K$^{\prime}$) couple with light and perform an interband transition only if the polarization of light coincides with the valley K (K$^{\prime}$) chirality.
This selection rule originates from the band topology of the Hamiltonian, reflecting the opposite Berry curvatures at K and K$^{\prime}$ and resulting in a disbalance of electron populations in the two valleys (and the anomalous Hall effect~\cite{Karch}).
The linear-response perturbation theory of light-matter coupling in TMDs has been developed in a number of works~\cite{gibertini2014spin, Li, Rostrami}. However, nonlinear optical phenomena~\cite{NLBoyd} remain largely unexplored~\cite{NLDirac1, NLDirac2}.

\begin{figure*}[!t]
	\includegraphics[width=0.80\linewidth]{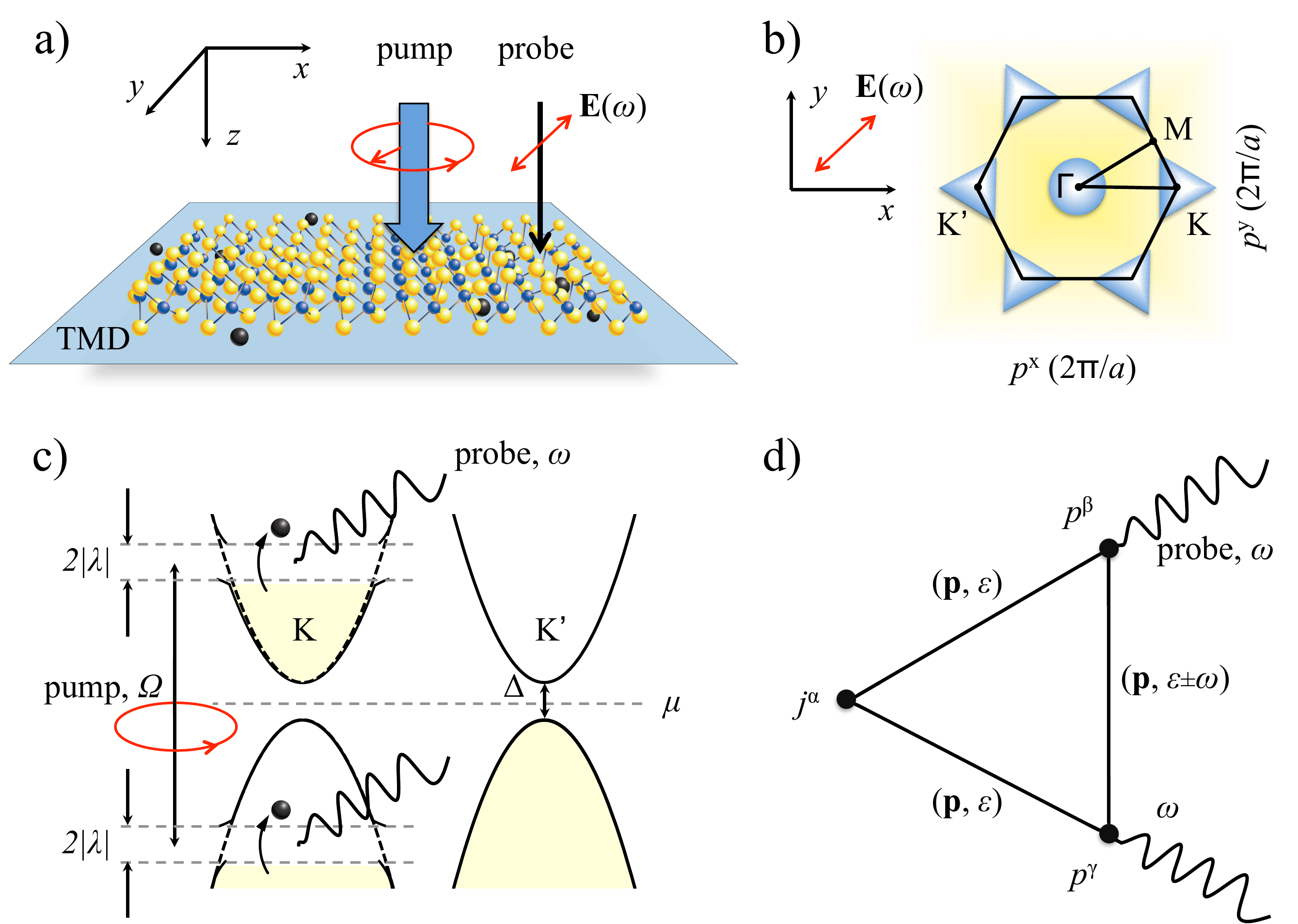}
	\caption{(a) System schematic: a transition metal dichalcogenide (TMD) monolayer exposed to the pump and probe EM fields. (b) The first Brillouin zone of the lattice. (c) K and K$^\prime$ valleys under the action of circularly polarized pump field with frequency $\Omega$; $2\lambda$ are the gaps. (d) The Feynman diagram(s), showing the method of calculation of the components $j^{\alpha}$ of PGE current.}
	\label{Fig1}
\end{figure*}

In this manuscript, we demonstrate that an unconventional PGE can occur in a 2D material if the valleys have different populations since the system is exposed to strong circularly-polarized light. The monolayer is initially non-conducting since the conductivity bands of both the valleys are empty. We expose the system to two EM fields. The first pumping circularly-polarized EM field with strong intensity populates one of the valleys, whereas the other valley remains empty. Due to the intravalley scattering, the valley reaches the saturation regime, when all the energy states in the valley are populated below some energy, which is determined by the EM field frequency. This regime is characterized by first, strong stationary nonequilibrium electron distribution function and second, strong modification of the energy spectrum of photogenerated electrons. The intraband dynamical gap opens, and its size is determined by the intensity of the pump EM field.
The second probe linearly-polarized EM field opens the intraband transitions, resulting in the uncompensated PGE current.

%--------------------------
%--------------------------
%--------------------------

\section{II. Theory}
The system schematic is presented in Fig.~\ref{Fig1}(a). To describe it, we start with the Hamiltonian in the electron representation:
\begin{equation}\label{eq1}
H_0 = \left(
        \begin{array}{cc}
          \frac{\Delta}{2}+\frac{\textbf{p}^2}{2m}+w_c(\textbf{p}) & 0 \\
          0 & -\frac{\Delta}{2}-\frac{\textbf{p}^2}{2m}-w_v(\textbf{p}) \\
        \end{array}
      \right),
\end{equation}
where the energy is counted from the middle of the TMD layer bandgap $\Delta$ and $\mathbf{p}$ is the momentum; we assume equal electron effective masses in the conduction and valence bands $m$, and we disregard the spin-orbit splitting of the valence band; $w_{c,v}(\textbf{p})=\eta C_{c,v}p^3\cos(\varphi_\textbf{p})\equiv\eta C_{c,v}(p_x^3-3p_xp^2_y)$ are the trigonal valley warping corrections to the electron dispersion in the corresponding bands [see Fig.~\ref{Fig1}(b)], $p=|\mathbf{p}|$, $\eta=\pm1$ is a valley index, and the parameters $C_{c,v}$ describe the strength of the warping.
%The choice of the latter equality corresponds to reflection symmetry around the $x$ axis.

The external EM fields acting on the monolayer are introduced by the Pierls' rule and the total Hamiltonian reads
\begin{gather}\label{eq2}
H = H_0+\left(
        \begin{array}{cc}
          0 & \lambda e^{-i\Omega t} \\
          \lambda^* e^{i\Omega t} & 0 \\
        \end{array}
      \right)
      +\frac{e}{mc}\textbf{pA}(t)\sigma_z,
\end{gather}
where $\lambda$ is the interband matrix element of the pump field with frequency $\Omega>\Delta$ [see Fig.~\ref{Fig1}(c)].
Within the parabolic band approximation, the matrix element possesses the following property: $|\lambda|^2\sim|1+\eta\sigma|^2I$, reflecting the valley selective optical interband transitions under the pump EM field with circular polarization $\sigma=\pm1$ and intensity $I$~\cite{OurNJP}; $\textbf{A}(t)$ is the vector potential of the probe EM field, $\sigma_z$ is a Pauli matrix.
%
%The probe EM field is described by the third term in Eq.~(\ref{eq2}). As we will show, It results in the intraband perturbation of the optically-pumped carriers and causes photogalvanic current in the second order with respect to $\textbf{A}(t)$.

The current density operator reads $\hat{\textbf{j}}=-e\partial \hat{H}_0/\partial \textbf{p}$, thus $j^\alpha=i\,\textmd{Sp}\left[\hat{j}^\alpha G^<(t,t)\right]$, where $G^<(t,t')$ is the lesser Green's function. Here and below we use the upper index to indicate the Cartesian components and the lower index to indicate the matrix elements.
Figure~\ref{Fig1}(d) shows the Feynman diagrams, which we use to find the current density.
The probe field is assumed to be weak, hence to calculate the PGE current we use the second-order response theory:
%The nonzero current occurs in the second order with respect to probe field
%
\begin{gather}\label{eq3}
j^\alpha(t)=\int_{C} dt'\int_{C} dt''\chi_{\alpha\beta\gamma}(t,t',t'')A^{\beta}(t')A^{\gamma}(t''),\\\nonumber
\chi_{\alpha\beta\gamma}(t,t',t'')=i\left(\frac{e}{mc}\right)^2\times~~~~~~~~~~~~~~~~~~~~~~~~~~~~\\\nonumber
~~~~~\times\,\textmd{Sp}\left[\hat{j}^\alpha G(t,t')p^\beta\sigma_z G(t',t'')p^\gamma\sigma_z G(t'',t)\right]_{C}.
\end{gather}
Here $C$ is the Keldysh contour.
%This expression is written as a perturbation theory for probe field and describe the system response to the probe EM field in the second order.
%
After some derivations, Eq.~(\ref{eq3}) gives two terms. The first one is time-independent, it describes the stationary PGE current. The second term contains the double frequency of the probe field, describing the second harmonic generation phenomena, which is beyond the scope of this manuscript.
%In this work, we will be interested in the first effect, thus disregarding the other.

The pump field should be taken into account in a nonperturbative manner. Thus the Green's functions in Eq.~(\ref{eq3}) depend on the times $t,~t'$ separately:
\begin{gather}\label{eq4}
G^{-1}(t,t')=\left(
        \begin{array}{cc}
          i\partial_t-\Delta_c(p) & -\lambda e^{-i\Omega t} \\
          -\lambda^* e^{i\Omega t} & i\partial_t+\Delta_v(p) \\
        \end{array}
      \right)
      \delta(t-t'),
\end{gather}
where for convenience we denoted $\Delta_{c,v}(p)=\frac{\Delta}{2}+\frac{\textbf{p}^2}{2m}+w_{c,v}(\textbf{p})$. The Green's funciton in Eq.~(\ref{eq4}) can be easily found using a unitary transformation to the rotating frame by the operator $S(t)=\exp(-i\sigma_z\Omega t)$, yielding:
\begin{gather}\label{eq5}
G(t,t')
=\left(
                          \begin{array}{cc}
                            g_{cc}(t-t')e^{-i\frac{\Omega}{2}(t-t')} & g_{cv}(t-t')e^{-i\frac{\Omega}{2}(t+t')} \\
                            g_{vc}(t-t')e^{i\frac{\Omega}{2}(t+t')} & g_{vv}(t-t')e^{i\frac{\Omega}{2}(t-t')} \\
                          \end{array}
                        \right),
\end{gather}
where
\begin{gather}\label{eq6}
g^{R,A}_{ij}(\textbf{p},\varepsilon)=\frac{\left(
                                             \begin{array}{cc}
                                               u_\textbf{p}^2 & u_\textbf{p}v_\textbf{p} \\
                                               u^*_\textbf{p}v^*_\textbf{p} & v_\textbf{p}^2 \\
                                             \end{array}
                                           \right)
}{\varepsilon-\varepsilon_1\pm i/2\tau}+\frac{\left(
                                             \begin{array}{cc}
                                               v_\textbf{p}^2 & -u_\textbf{p}v_\textbf{p} \\
                                               -u^*_\textbf{p}v^*_\textbf{p} & u_\textbf{p}^2 \\
                                             \end{array}
                                           \right)
}{\varepsilon-\varepsilon_2\pm i/2\tau},
\end{gather}
and
\begin{gather}\label{eq7}
\left(
  \begin{array}{c}
    u_\textbf{p}^2 \\
    v_\textbf{p}^2 \\
  \end{array}
\right)=\frac{1}{2}\Biggl[1\pm\frac{\xi+\frac{w_c+w_v}{2}}{\sqrt{\left(\xi+\frac{w_c+w_v}{2}\right)^2+|\lambda|^2}}\Biggr],
\\
\nonumber
       \varepsilon_{1,2}=\frac{w_c-w_v}{2}\pm\sqrt{\left(\xi+\frac{w_c+w_v}{2}\right)^2+|\lambda|^2},\\\nonumber
       \xi=\frac{\textbf{p}^2}{2m}-\frac{\Omega-\Delta}{2}.
\end{gather}
Here $\varepsilon_{1,2}$ are dispersions of quasiparticles in the presence of resonant pumping EM field, and $\tau$ is a momentum relaxation time.
%
%From Eq.~(\ref{eq7}) it follows that the pump field opens a dynamical gap $|\lambda|$ in the dispersions $\varepsilon_{1,2}$ of photon-dressed quasiparticles, see Fig.~\ref{Fig1}(c).

Using the linearly-polarized probe field $\textbf{A}(t)=\textbf{A}_0\exp(-i\omega t)/2+$c.c. and applying the Keldysh diagrammatic technique~\cite{Keldysh}, we find
\begin{gather}\label{eq8}
\chi_{\alpha\beta\gamma}(\omega)=i\left(\frac{e}{2mc}\right)^2\sum_{\textbf{p};i=c,v}j^\alpha_{ii}p^\beta p^\gamma\left[F_i(\textbf{p},\omega)+F_i(\textbf{p},-\omega)\right],\\\nonumber
F_i(\textbf{p},\omega)=\sum_\varepsilon(n_\varepsilon-n_{\varepsilon-\omega})g^R_{ii}(\textbf{p},\varepsilon)g^A_{ii}(\textbf{p},\varepsilon)\times\\\nonumber
\times\Biggl[g_{ii}^R(\textbf{p},\varepsilon-\omega)
-g^A_{ii}(\textbf{p},\varepsilon-\omega)\Biggr],
\end{gather}
where $n_\varepsilon$ is the nonequilibrium quasiparticle distribution function (as opposed to equilibrium electron distribution).
In general case it depends on the intensity of the pump EM field, intraband relaxation, interband recombination, and intervalley scattering times~\cite{OurNJP}.
As it follows from Eq.~(\ref{eq8}), the contributions of electrons from the conduction and valence bands have similar structure and they sum up. Thus, we can consider only one of the bands and after extend the results on the other band.

%------------------------
%------------------------
%------------------------

\section{III. Results and discussion}

The action of the pump field not only results in the population of the valley, but also opens a dynamical gap $2|\lambda|$ in quasiparticle dispersion, see Eq.~(\ref{eq7}) and Fig.~\ref{Fig1}(c). Obviously, the PGE current at $T=0$ occurs only if $\omega>2|\lambda|$. We consider here quasiballistic electron motion, assuming that the scattering is weak enough (or the intensity of the pump field is strong enough), so that $|\lambda|\tau\gg1$.

Combining together Eqs.~(\ref{eq6}), (\ref{eq7}) and~(\ref{eq8}), the conduction band contribution to the PGE tensor $\chi_{\alpha\beta\gamma}$ reads
\begin{eqnarray}\label{eq9}
\chi^c_{\alpha\beta\gamma}(\omega)&=&\pi\tau\left(\frac{e}{2mc}\right)^2\int\frac{d\textbf{p}}{(2\pi)^2}j^\alpha_cp^\beta p^\gamma
\times\\\nonumber
&~~\times& u^2_\textbf{p}v^2_\textbf{p}(u^2_\textbf{p}-v^2_\textbf{p})(n_{\varepsilon_1}-n_{\varepsilon_1-\omega})\delta(\varepsilon_1-\varepsilon_2-\omega).
\end{eqnarray}
The valence band contribution can be found from Eq.~(\ref{eq9}) by the replacement $j^\alpha_c\leftrightarrow j^\alpha_v,\,\,u^2_\textbf{p}\leftrightarrow v^2_\textbf{p}$.

The intraband kinetics of photogenerated electrons under the action of resonant pump field has been considered in Refs.~\onlinecite{OurNJP, elesin1971coherent, Galitskii}. It was shown that if the intraband relaxation time is much smaller than the interband recombination time (saturation regime), then the distribution function of the quasiparticles with the dispersions~(\ref{eq7}) has the form of the Fermi distribution with zero Fermi energy~\cite{Remark_Elesin}, $n_{\varepsilon}=(e^{\varepsilon/T}+1)^{-1}$.
%
%\begin{gather}\label{eq10}
%\end{gather}
%
In the limit of zero temperature, $(n_{\varepsilon_1}-n_{\varepsilon_1-\omega})\rightarrow-\theta[\omega-\varepsilon_1]$.

As it has been pointed out above, the PGE comes from the trigonal warping of the electron spectrum. Due to the smallness of this effect, we expand the current up to the linear-order corrections in $w_c$ and $w_v$. These warping terms are contained in the  $u_\textbf{p},~v_\textbf{p}$ coefficients and in the quasiparticle dispersions $\varepsilon_1,\,\varepsilon_2$. Expanding Eq.~\eqref{eq9} and combining the contributions from the conduction and valence bands, we find
\begin{gather}\label{eq10}
    \chi_{\alpha\beta\gamma}(\omega)
=\frac{2e^3\pi\tau}{4m^2c^2}\int\frac{d\textbf{p}}{(2\pi)^2}
\Bigl[\frac{p^\alpha}{m}\left(\frac{w_c+w_v}{2}\right)\frac{dP(\xi)}{d\xi}\\
\nonumber
+P(\xi)\frac{\partial}{\partial p^\alpha}\left(\frac{w_c+w_v}{2}\right)\Bigr]
p^\beta p^\gamma,
\end{gather}
where
\begin{gather}\label{eq10.1}
P(\xi)=u^2_\textbf{p}v^2_\textbf{p}(u^2_\textbf{p}-v^2_\textbf{p})\delta(\varepsilon_1-\varepsilon_2-\omega)\Bigr|_{w_c+w_v=0}\\
\nonumber
=\frac{|\lambda|^2\xi}{4\varepsilon_\textbf{p}^3}\delta(2\varepsilon_\textbf{p}-\omega).
\end{gather}
Here $\varepsilon_\textbf{p}=\sqrt{\xi^2+|\lambda|^2}$ is the quasiparticle dispersion in the absence of warping. If the frequency of the pump field satisfies $|\lambda|\ll(\Omega-\Delta)/2$, then instead of the integration over the momentum in~(\ref{eq10.1}) we can perform the $\xi$-integration, replacing
$$\int pdp\rightarrow m\int\limits_{-\infty}^{\infty}d\xi.$$

The analysis of Eq.~(\ref{eq10}) shows that the nonzero elements of PGE tensor read
\begin{gather}\label{eq11}
\chi_{yxy}(\omega)=\chi_{yyx}(\omega)=\chi_{xyy}(\omega)=-\chi_{xxx}(\omega)\neq 0,
\end{gather}
that allows us to find all the components, calculating only the $\chi_{xxx}(\omega)$ component.
Performing the integration in Eq.~(\ref{eq10}), we find (restoring the Plank constant):
\begin{gather}\label{eq11}
\chi_{xxx}(\zeta)=\chi_0\frac{\sqrt{\zeta^2-1}}{\zeta^2}\theta[\zeta^2-1],\,\,\,\zeta=\frac{\omega}{2|\lambda|},\\\nonumber
\chi_0=3\eta\left(\frac{C_c+C_v}{2}\right)\frac{em^2|\lambda|\tau}{2\hbar^3}\left(\frac{ep_0}{2mc}\right)^2,
\end{gather}
where $p_0=\sqrt{m(\Omega-\Delta)}$. We see that the current is proportional to  $\eta|\lambda|\propto\eta|1+\eta\sigma|$, which determines the sensitivity of the current to valley quantum number, polarization of the pump EM field, and the factor $|\lambda|\tau\gg1$.

It should be noted, that here we derived the expression for the PGE tensor in the case when the TMD layer is initially in the dielectic regime with the chemical potential lying in the bandgap. The generalization for the case of n- or p-doped TMDs can be done by replacing $\Delta$ with its value shifted by the Fermi energy.
We have disregarded here the possible spin splitting of the bands. The spin quantum number should be conserved in the interband optical transitions,  and the respective contributions to the PGE current are just summed up.
\begin{figure}[!t]
	\includegraphics[width=1\linewidth]{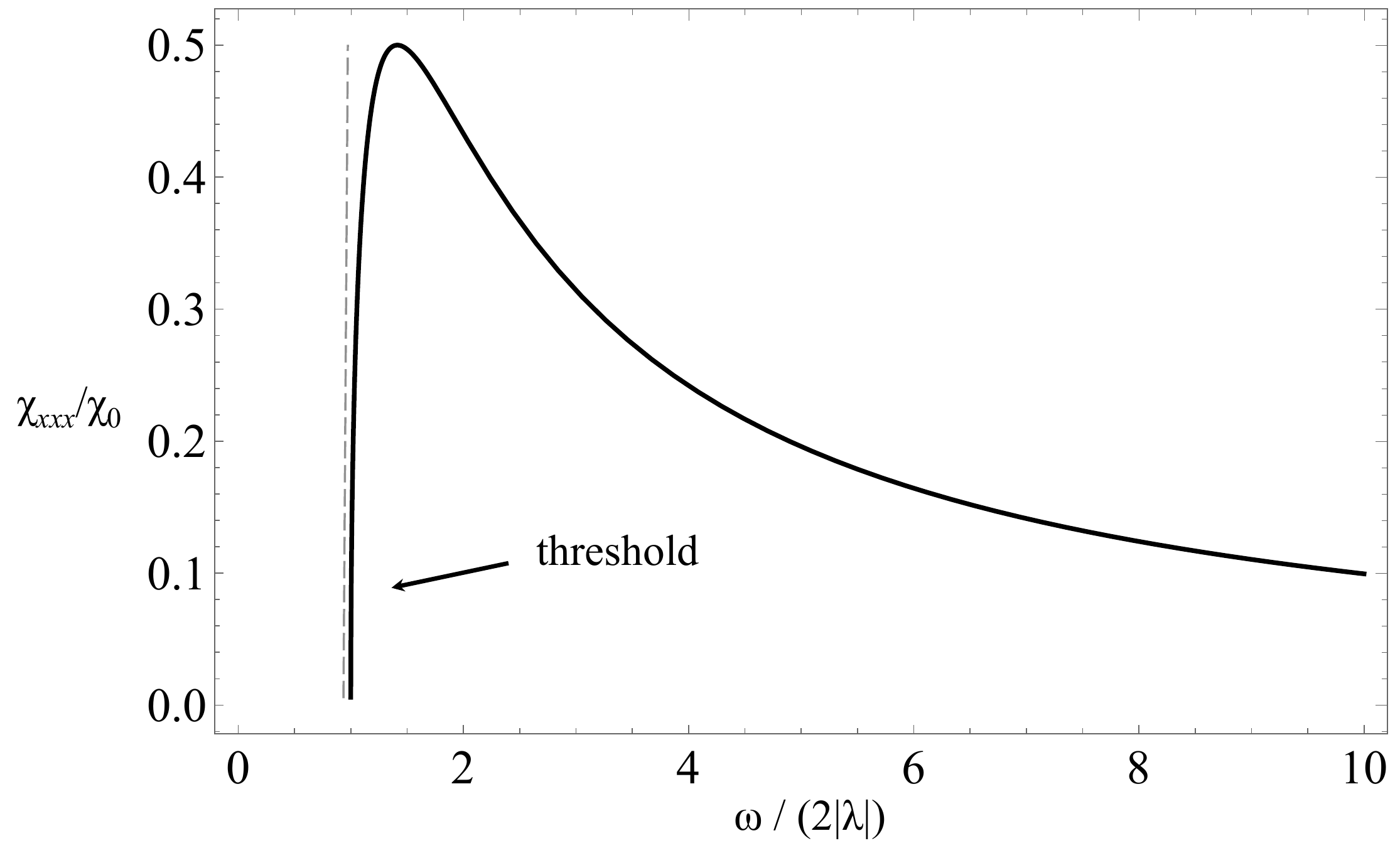}
	\caption{Normalized spectrum of the photogalvanic tensor~\eqref{eq11} in the vicinity of the threshold (marked by the grey dashed line).}
	\label{Fig2}
\end{figure}

Figure~\ref{Fig2} shows the spectrum of $\chi_{xxx}(\zeta)$. The qualitative explanation of such dependence of the current on the frequency is the following. In the vicinity of the dynamical-induced gap, the density of states (DOS) is renormalized. Indeed, using the standard formula
\begin{eqnarray}\label{eq8.1}
\rho(\epsilon)=-\frac{1}{\pi}\textmd{Im}\,\sum_{\textbf{p}}g^R_{cc}(\textbf{p},\epsilon),
\end{eqnarray}
we find
\begin{eqnarray}\label{eq8.3}
\rho(\epsilon)=\rho_0\frac{\epsilon}{\sqrt{\epsilon^2-|\lambda|^2}}\theta[|\epsilon|-|\lambda|],
\end{eqnarray}
where $\rho_0=m/2\pi$ is the DOS of the 2D system in the absence of a pump. We see that DOS~\eqref{eq8.3} drastically increases in the vicinity of dynamical bandgap $|\lambda|$. At the same time, the velocity of quasiparticles $\partial_\textbf{p}\varepsilon_\textbf{p}=\textbf{p}\xi/(m\varepsilon_\textbf{p})$ is zero at $p=p_0$, suppressing the PGE current at the threshold. Thus, the quasi-resonant behavior of PGE current in Fig.~\ref{Fig2} is the combined effect of these two factors.

%--------------------
%--------------------
%--------------------

\section*{Conclusions}
We have developed a microscopic quantum theory of an unconventional photogalvanic effect in 2D Dirac semiconductors under the action of strong pumping electromagnetic field. 
We have demonstrated, that the emergence of photon-dressed quasiparticles and the dynamical gap opening result in a thresholdlike behavior of the current as a function of the probe field frequency due to the dynamical renormalization of the density of states and quasiparticle velocity. 

Our results can be extended to other materials, possessing a similar band structure as TMDs and obeying the valley-dependent interband optical selection rules. Moreover, the appearance of dynamically-induced gaps and the renormalization of the density of states open a way for engineering dispersions of quasiparticles in order to affect valley-selective   second-order response effects, such as the photon drag effect and, possibly, the second harmonic generation.

%--------------------------
%--------------------------
%--------------------------

\section*{Acknowledgments}
We acknowledge stimulating discussions with M.~Entin. We also thank E.~Savenko for the help with the figures. This research has been supported by the Russian Science Foundation (Project No.~17-12-01039) and the Institute for Basic Science in Korea (Project No.~IBS-R024-D1).

%----------------------------
%----------------------------
%----------------------------

%------------------------
%------------------------
%------------------------

\end{document}